\documentclass{article}
\usepackage{graphicx}
\usepackage{PRIMEarxiv}
\usepackage{amsmath}
\usepackage[linesnumbered,lined,commentsnumbered, ruled]{algorithm2e}
\usepackage{float}
\title{Distance based prefetching algorithms for mining of the sporadic requests associations}

\author{
  Vadim Voevodkin \\
  YADRO \\
  Higher School of Economics \\
  Nizhny Novgorod\\
  \texttt{v.voevodkin@yadro.com} \\
   \And
  Andrey Sokolov \\
  YADRO \\
  Moscow State University \\
  Moscow\\
  \texttt{andrey.sokolov@yadro.com} \\
}

\begin{document}
\maketitle

% \author[1]{Vadim Voevodkin}
% \author[1, 2]{Andrey Sokolov}
% \affil[1]{YADRO}
% \affil[2]{Moscow State University}
% \corrauthor[1]{Vadim Voevodkin}{v.voevodkin@yadro.com}

\begin{abstract}

% Abstract
Modern storage systems intensively utilize data prefetching algorithms while processing sequences of the read requests. Performance of the prefetchingalgorithm (for instance increase of the cache hit ratio of the cache system – CHR) directly affects overall performance characteristics of the storage system (read latency, IOPS, etc.).

There are widely known prefetching algorithms that are focused on the discovery of the sequential patterns in the stream of requests. 
This study examines a family of prefetching algorithms that is focused on mining of the pseudo random (sporadic) patterns between read requests – sporadic prefetching algorithms. The key contribution of this paper is that it discovers a new, lightweight family of distance-based sporadic prefetching algorithms (DBSP) that outperforms the best previously known results on MSR traces collection.Another important contribution of this paper is a thorough description of the procedure for comparing the performance of sporadic prefetchers.
\end{abstract}

% % keywords can be removed
% \keywords{
% Sporadic prefetching algorithms \and DBSP  \and }

\flushbottom
\maketitle
\thispagestyle{empty}

\section*{Introduction}

Data storage systems play a crucial role in ensuring efficient and timely retrieval of data in various applications. With the ever-increasing volumes of data, it has become imperative to improve the performance of these systems by reducing  access latency. 
Prefetching algorithms play a crucial role in modern cache systems, which are used in storage systems, CPUs etc. 
In this paper, we propose a new family of prefetching algorithms that is focused on mining of the pseudo random (sporadic) patterns between read requests – sporadic prefetching algorithms. The key contribution of this work is that it discovers a new lightweight family of distance-based sporadic prefetching algorithms (DBSP).

The storage system is composed of the components shown in Figure 1: host, storage backend, storage controller, cache and an optional component — prefetcher.

The host sends read requests to the storage controller. Storage controller asks cache systems if the  corresponding data is contained in it. If yes, then such event is treated as a cache hit, data is retrieved from the fast cache system and then delivered to the host. If the cache doesn't contain requested data, then this event is treated as a cache miss and the request is sent further to the storage backend. Storage backend usually contains a lot of HDDs and/or SSDs organized into RAID arrays. Thus, it performs much slower than the fast cache system, which usually uses RAM memory to store data. 

If the storage system contains a dedicated read prefetcher component, then additional processing happens when a read request arrives to the storage controller. After request arrival, the storage controller generates a corresponding event message and sends it to the prefetcher component.  

The purpose of the prefetcher is to predict which requests will come to the storage system in the nearest future after some specific initial read request. These requests are called requests associated with the initial read request or simply "associations". For different initial requests, prefetcher may return different sets of associations.
\begin{figure}[H]
	\begin{center}
		\includegraphics[scale=0.35
  ]{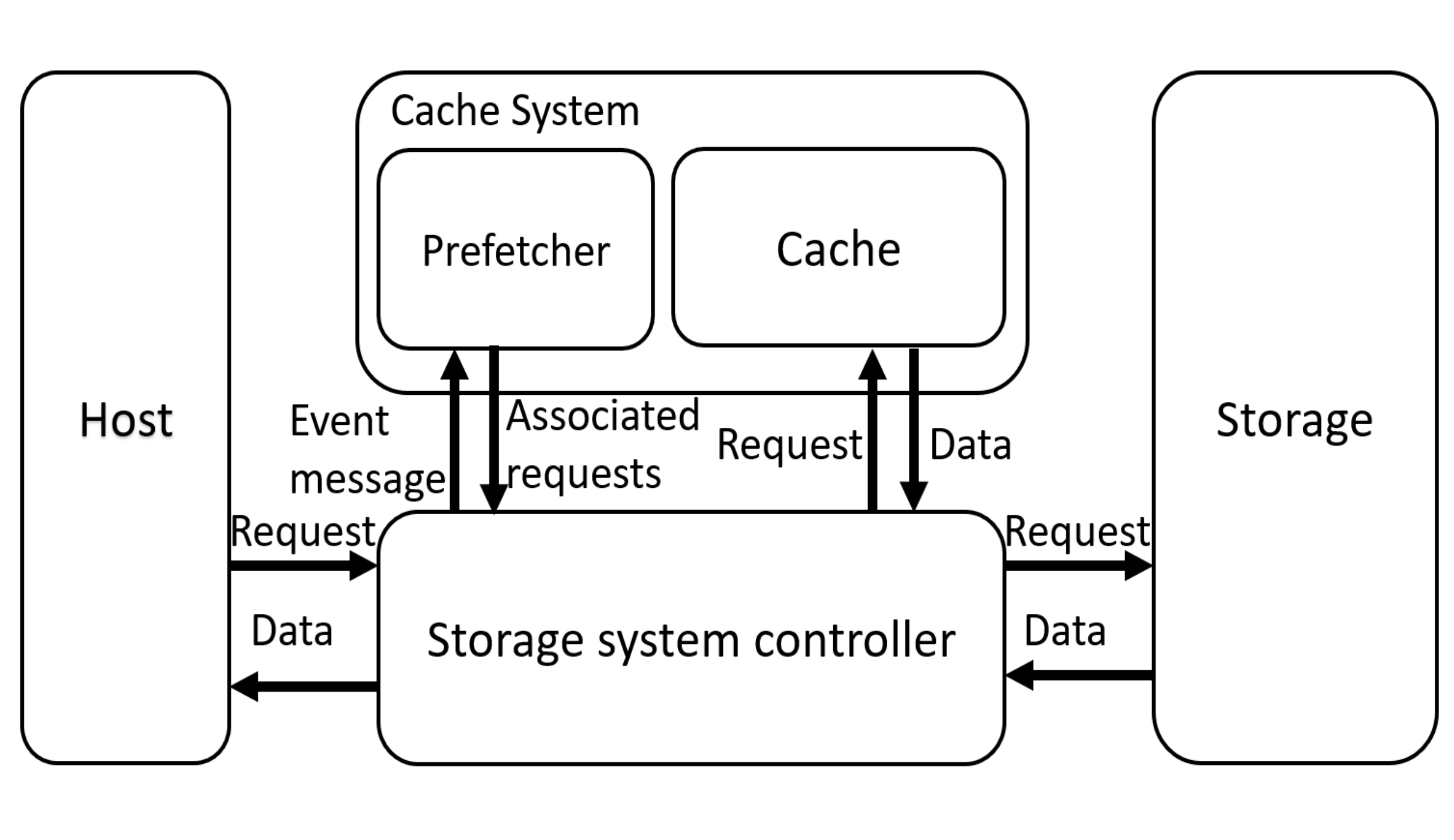}
	\end{center}
	\caption{ Scheme of work storage system controller }
	\label{pic:SAN}
\end{figure}

After receiving the initial read request, prefetcher tells about it's  associations to the storage controller before their actual arrival. Thus, storage controller is able to read corresponding data from storage backend in advance and put the data in the cache. 

Retrieving data from the cache is much faster than retrieving same amount of data from the storage backend. For example, read latency for storage disks typically ranges from 1 ms (SSDs) to 15 ms (HDDs), while read latency of the cache that is based on RAM is about 10 ns [1, 2, 3]. So if the prefetcher correctly predicts the future read request, then corresponding data will be retrieved to the user from the cache with the lowest possible latency. Thus correct predictions of the prefetcher improve overall performance characteristics of the storage system. In contrast – any incorrect prediction of the future request will result in the pollution of the cache. 

The key performance characteristics of data storage systems are latency and IOPS. The latency is the total delay between the arrival of the input read/write request and receiving data requested by the host. 
IOPS value is the number of Input/Output operations per second. 
The Storage controller communicates event messages about internal events and has the ability to process the prefetcher's predictions.  The controller in event messages reports incoming requests and asks the prefetcher for the associations.  The rules of how to use associations are overseen by the prefetching policy. This policy defines when it makes sense to prefetch associated requests from the storage backend. Decision may depend on such factors as request size, total load of the storage system and others. Prefetching  policies are not covered in this paper. For further investigation we use so called "Always" prefetch policy which means that all associations are being prefetched from the storage backend.  This allows us to find out the upper bound of the prefetching algorithm.

Comparing prefetching algorithms with each other is a non-trivial task.Various methodologies for comparison of the prefechers were proposed in other works [1, 2, 3, 4].  Unfortunately, comparison methodologies provided in these works lack many important details. For example, which datasets and algorithm parameters were used and other important particulars. In this paper, we will try to eliminate this drawback and present a detailed description of the methodology of comparing sporadic prefetching algorithms.

One of the de-facto standard caching algorithms is Least Recently Used (LRU) cache [3, 5].  
We will consider LRU cache as a basic caching algorithm. And, further, we will compare sole LRU cache with it's combinations with various prefetching algorithms.

There are several common and efficient prefetching algorithms that are focused on the  processing of the sequential read requests, such as AMP [1] or PG [2]. In this paper, we do not compare our novel prefetching algorithm (DBSP) with these prefetching algorithms because they belong to another family - sequential pattern prefetchers.

An appropriate algorithm was found for comparative analysis, which also targets sporadic data streams.
Mithril [4] is currently one of the best known prefetching algorithms focused on the sporadic read request sequences. We consider this algorithm as a main baseline for comparison.

% main result
The key contributions of the paper are: 
\begin{itemize}
    \item Firstly, this paper introduces the new family of read prefetching algorithms – Distance Based Sporadic Prefetcher (DBSP). These algorithms after optimization of its parameters on MSR traces collection succeeds to overcome Mithtil algorithm near  2\% in terms of CHR and  3\% in terms of the Precision. 
    \item The second, in this paper, we introduce the detailed methodology for evaluating the quality of the prefetching algorithms. 

\end{itemize}

Next we will give the detailed description of the novel developed prefetching algorithm, its asymptotic complexity will be analyzed and compared to the complexity of the Mithril.
After that, the methodology for evaluating the quality of prefetching algorithms will be provided.
Finally, we will present experimental results of DBSP algorithm performance.

\section{Distance Based Sporadic Prefetcher }
DBSP is prefetching algorithm. According to Figure 1, prefetcher is located next to the cache. For the DBSP to work, we allocate for it a part of the total amount of fast memory (in particular RAM) which makes it fair to compare it's characterizes with the sole LRU cache of the same size. To further explain the work of the algorithm, let us introduce  hyperparameters and general structures of the DBSP algorithm. This article defines a specific table structure as the combination of a hash table  and a doubly-linked list. DBSP uses three such tables: record table, compute table and prefetch table.  Record table (rTable) and  Compute table (cTable) are used for accumulate history of insert read requests.  Prefetch table (pTable) is used to contain any specific structure which is described later.   
The following are the  hyperparameters of the DBSP algorithm:
 \begin{itemize}
      \item $nr_r$, $nr_c$, $nr_p$ - number of rows in record, compute and prefetch tables correspondingly.
     \item $l_a$  is some confidence interval to compute associations.
     \item $s_{c}$ is the maximum number of associated requests in prefetch table.
     \item $l_{min}$, $l_{max}$ - minimal and  maximal number of $ts$ in each rows of compute table correspondingly.
 \end{itemize}

The DBSP prefetching algorithm is described below (Algorithm 1). The main purpose of this prefetch  algorithm is to accumulate the history of arrived requests by periodic calling of the function -  $prefetch\_engine$ (see Algorithm 1) and return associated requests if it is necessary. The $prefetch\_engine$  is used to accumulate history of insert read requests in all specific tables.

Denote $r$ - read request, which is got from storage system controller. Each request consists of a unique address and size in blocks. All internal requests are contained in the record table. 
 The concept of a \textit{right request } must be introduced. Assume that 
$\mathrm{h_r} = \left(ts_1,...,ts_N\right)$ - history of it's arrival times of the request $r$ and $h_a(i)$ - i-s timestamp from $\mathrm{h_r}$. Timestamps $ts_i$ are usually measured by the operating system. So if   $len(\mathrm{h_r})$ is greater than the parameter, $l_{min}$ we will consider request as right.
If the request is classified as right, it is moved to the compute table.
 Numerical experiments showed that reliable detection of associations needs it's at least two times repetition of arriving request ( $l_{min} = 2$). When the compute table is completely filled, the association search algorithm is started to find associated requests  (see \newline Algorithm 2). 
 This algorithm  is moving associated requests into prefetch table.

Let us define a function for association degree for two request histories:

\text{where $f(h_a, h_b)$ is the distance between histories of  the two requests $a$ and $b$.}
It is important to note that the association degree is not symmetric.    
   Request is  moved into   prefetch table, if it's association degree is bigger than degrees of another pair of requests from prefetch table.

 In this article, we used  following distance functions:

 $$f_1(h_a, h_b) = \sum_{i = 1 }^{len(h_a)}(|h_a(i) - h_b(i)|) $$
 $$f_2(h_a, h_b) = \frac{1}{len(h_a)}\sum_{i = 1 }^{len(h_a)}(|h_a(i) - h_b(i)|) $$

 note that first one is the Manhattan distance and the second one is its normalized version. Implementation of moving associated request from compute table to prefetch table  is presented in (Algorithm 2).

Consider the time complexity of adding new queries to the record table.
The asymptotic complexity time of adding new arrived requests  into Record and Compute table  is average linear $O(n)$, where $n$ is the number of input read requests from storage system controller. Since   the insert and remove operation in  hash table   has an average constant  complexity, while  doubly-linked list supports constant time insertion and removal of elements from anywhere in the list.   

 The asymptotic complexity time of (Algorithm 2) is  $O(l_a k )$, where $k$ is a num of rows in compute table. This function iterate through all rows compute table. Each request from compute table is compared with $l_a$ next rows. The most similar  pairs are added to the prefetch table. 

\RestyleAlgo{plain}
\IncMargin{1em}
\begin{figure}[H]
    \hrule
    \vspace{2pt}
        \textbf{Algorithm 1:} DBSP
    \vspace{2pt}
    \hrule

\begin{minipage}[t]{0.48\linewidth} 
\IncMargin{1em}
\begin{algorithm}[H]
\SetKwData{Left}{left}\SetKwData{This}{this}\SetKwData{Up}{up}
\SetKwInOut{Input}{input}\SetKwInOut{Output}{output}
\SetAlgoLined\SetArgSty{}

% \Output{...}
\BlankLine
 \label{Algo:part1}
        \SetAlgoLined\SetArgSty{}
        \BlankLine
        \SetKwProg{RW}{DBSP}{:}{\KwOut{\textit{Associated container} a\_c}}
        \RW {(record table \textit{rTable}, compute  table \textit{cTable},
prefetch table \textit{pTable}, min support $t_{min}$,
request \textit{r}, time stamp \textit{ts}, prefetch\_flag \textit{pref\_flag}, normalized measure \textit{normalised})}
{

prefetch\_engine(\textit{rTable}, \textit{cTable}, \textit{pTable},  $t_{min}$, \textit{r}, \textit{ts}, \textit{pref\_flag}, \textit{normalised} );

\eIf{pref\_flag}{
a\_c := pTable.find(r);\\

}{
a\_c :=  None;
}}

\label{Algo:insert in cont}
\end{algorithm}\DecMargin{1em}
\end{minipage} 
\hfill
\begin{minipage}[t]{0.48\linewidth} 
\IncMargin{1em}
\begin{algorithm}[H]
\SetKwData{Left}{left}\SetKwData{This}{this}\SetKwData{Up}{up}
\SetKwInOut{Input}{input}\SetKwInOut{Update}{Update}
\SetAlgoLined\SetArgSty{}

% \Output{...}
\BlankLine
 \label{Algo:part2}
        \SetAlgoLined\SetArgSty{}
        \BlankLine
        \SetKwProg{RW}{prefetch\_engine}{:}{\Update{\textit{rTable}, \textit{cTable}, \textit{pTable}}}
        \RW {(record table \textit{rTable}, compute  table \textit{\textit{cTable}},
prefetch table \textit{pTable}, min support $t_{min}$,
request \textit{r}, time stamp \textit{ts}, normalized measure \textit{normalised})}
{

\eIf{\textit{rTable}.find(r)}{\textit{rTable}[r].push\_back(ts)
\\
\If{len(\textit{rTable}[r]) == $t_{min}$}{\textit{\textit{cTable}}[r] = \textit{rTable}[r]\\ 
\textit{rTable}.remove(r)}}{
\eIf{len(\textit{\textit{cTable}}[r]) == $t_{max}$} {\textit{\textit{cTable}}.remove(r)}{\textit{\textit{cTable}}[r].push\_back(ts)}
}

\If{\textit{\textit{cTable}}.is\_full()}{compute\_associated\_requests( \textit{cTable},   
 \textit{pTable}, \textit{normalised}, \textit{s\_c}, \textit{ l\_a})}
}

\label{Algo:insert in containe}
\end{algorithm}\DecMargin{1em}
\end{minipage} 
\hrule
\end{figure}
\DecMargin{1em}
\RestyleAlgo{ruled}
\setcounter{algocf}{1}

\RestyleAlgo{plain}
\IncMargin{1em}
\begin{figure}[H]
    \hrule
    \vspace{2pt}
    
        \textbf{Algorithm 2:} get requests association degrees
    \vspace{2pt}
    \hrule

\begin{minipage}[t]{0.48\linewidth} 
\IncMargin{1em}
\begin{algorithm}[H]
\SetKwData{Left}{left}\SetKwData{This}{this}\SetKwData{Up}{up}
\SetKwInOut{Input}{input}\SetKwInOut{Update}{Update}
\SetAlgoLined\SetArgSty{}

% \Output{...}
\BlankLine
 \label{Algo:computing__}
        \SetAlgoLined\SetArgSty{}
        \BlankLine
        \SetKwProg{RW}{compute\_associated\_requests}{:}{\Update{\textit{pTable}}}
        \RW {(compute  table \textit{cTable}, 
prefetch table \textit{pTable}, normalized measure \textit{normalised}, container size \textit{s\_c}, lookahead \textit{ l\_a})}
{
\For {$i$ = 0 to $len(\textit{\textit{cTable}})$}{
    \For {$j$ = 0 to $l\_a$}{
        cur\_degree := get\_association\_degree(\textit{\textit{cTable}}[i], \textit{\textit{cTable}}[j], normalised)\\
        \textit{pTable}[\textit{\textit{cTable}}[i]].insert\_in\_container(\\ \textit{\textit{cTable}}[j], cur\_degree, $s\_c$)
    }
}}

\label{Algo:insert in container}
\end{algorithm}\DecMargin{1em}
\end{minipage} 
\hfill
\begin{minipage}[t]{0.48\linewidth} 
\IncMargin{1em}
\begin{algorithm}[H]
\SetKwData{Left}{left}\SetKwData{This}{this}\SetKwData{Up}{up}
\SetKwInOut{Input}{input}\SetKwInOut{Update}{Update}
\SetAlgoLined\SetArgSty{}

% \Output{...}
\BlankLine
\label{Algo:computing_degree}
        \SetAlgoLined\SetArgSty{}
        \BlankLine
        \SetKwProg{RW}{get\_association\_degree}{:}{\KwOut{$degree$}}
        \RW {(vector of timestamps $h_a$, vector of timestamps $h_b$, normalized measure \textit{normalised})}
{

\If{len($h_a$) $\neq$ len($h_b$) or $h_a$[0] \textgreater $h_b$[0]}{
return 0}
$degree := 0$

\For {
$iter = 0$ 
to len($h_a$)}{
    $degree$ += abs ($h_a$[i] - $h_b$[i])\\
               }

\If{normalised}{ degree = (degree / len($h_a$))}
$degree = degree^{-1}$}

\end{algorithm}\DecMargin{1em}
\end{minipage} 
\hrule
\end{figure}
\DecMargin{1em}
\RestyleAlgo{ruled}
\setcounter{algocf}{2}

% Consider the time complexity of adding new queries to the record table.
% The asymptotic complexity time of adding new arrived requests  into Record and Compute table  is average linear $O(n)$, where $n$ is the number of input read requests from storage system controller. Since   the insert and remove operation in  hash table   has an average constant  complexity, while  doubly-linked list supports constant time insertion and removal of elements from anywhere in the list.   

%  The asymptotic complexity time of (Algorithm 2) is  $O(l_a k )$, where $k$ is a num of rows in compute table. This function iterate through all rows compute table. Each request from compute table is compared with $l_a$ next rows. The most similar  pairs are added to the prefetch table. 

For further analysis of the computational complexity of working with a prefetch table, let us introduce the concept of associative container and associated requests. A pair of right requests  is  associated (associated requests) if the degree of closeness is not equal to $+\infty$. Associative container is a specific data structure for collecting associated requests. This container  consists of a hash table and a binary tree. The hash table is used to quickly check the presence of such an association in the container, and the binary tree is used for the ordered storage of associations and simple maintenance of the ordered tree structure. Algorithm 3 presents an implementation of the insert operation. 

The prefetch table stores associations in a container, allowing associations to be added and removed in $ O(n)$ time on average, where $n$ is the number of input associated requsts from function \textit{compute\_associated\_requests}. Additionally, it is important to provide an estimate of the time complexity involved in maintaining the associated container structure. Insert, remove and search  operations have an average complexity $O(\log(s_c))$.

\begin{minipage}[t]{\linewidth} 

\begin{minipage}[t]{0.48\linewidth} 
\IncMargin{1em}
\begin{algorithm}[H]
\SetKwData{Left}{left}\SetKwData{This}{this}\SetKwData{Up}{up}
\SetKwInOut{Input}{input}\SetKwInOut{Update}{Update}
\SetAlgoLined\SetArgSty{}

% \Output{...}
\BlankLine
 \label{Algo:computing}
        \SetAlgoLined\SetArgSty{}
        \BlankLine
        \SetKwProg{RW}{insert\_in\_container}{:}{\Update{\textit{pTable}}}
        \RW {( request \textit{r}, degree of association \textit{degree}  ,  container
size $s\_c$)}
{

\eIf{check r in container}
{container.remove(r);\\ container.insert(r);}
{container.insert(r);}

\If{container.size() \textgreater $s\_c$}{
min\_degree\_conrainer := container.min\_degree();\\
container.erase(min\_degree\_conrainer);
}
}

\caption{insert in container}
\label{Algo:insert in contai}
\end{algorithm}\DecMargin{1em}
\end{minipage}
\hfill
\begin{minipage}[t]{0.48\linewidth} 

 \IncMargin{1em}
\begin{algorithm}[H]
\SetKwData{Left}{left}\SetKwData{This}{this}\SetKwData{Up}{up}
\SetKwInOut{Input}{input}\SetKwInOut{Output}{output}\SetKwInOut{Update}{Update}
\SetAlgoLined\SetArgSty{}

\BlankLine
 \label{Algo:computing_}
        \SetAlgoLined\SetArgSty{}
        \BlankLine
        \SetKwProg{RW}{evaluate\_prefetcher}{:}{\Update{avg. CHR, avg. SAR, avg. precision}}
        \RW {(prefetcher \textit{pref}, caching algorithm \textit{cache},  set\_traces \textit{traces}, prefetcher relative size \textit{pref\_rel\_size},cache size \textit{ cache\_size}, storage size \textit{storage\_size})}
{

 \textit{chr, sar, precision} := [], [], [];
 
\For{trace in \textit{traces}}{
  cur\_chr, cur\_sar, cur\_presicion := get\_target\_metrics( \textit{pref},  \textit{cache},   \textit{trace},  \textit{pref\_rel\_size}, \textit{cache\_size},  \textit{storage\_size});
  
  \textit{chr}.append(\textit{cur\_chr});
  
  \textit{sar}.append(\textit{cur\_sar});
  
  \textit{precision}.append(\textit{cur\_presicion});
}
}

\caption{evaluation methodology}
\label{Algo:compute_t}
\end{algorithm}\DecMargin{1em}

\end{minipage}
\end{minipage}
\section{Prefetcher Evaluation Methodology} 
Prefetcher evaluation methodology is an algorithm for comparing different types of prefetchers.  This algorithm   maps a specific prefetching algorithm into  a set of  metrics.
This algorithm (see Algorithm 4)  takes as input a caching algorithm, a prefetching algorithm, a set of traces and system parameters.
Thus, using this methodology, it is possible to study and compare any pair of prefetching  and caching algorithms that work in the system with storage and cache.  Therefore, comparisons can be made not only for storage systems, but also for example in CPU cache. The list of evaluated metrics consists of \textit{cache hit ratio (CHR), precision } and \textit{storage activity ratio (SAR)}. All of them will be described below. 

\textbf{Cache Hit Ratio (CHR).}
Cache hit rate is one of the main characteristics of the prefetching algorithms. It's calculated as follows: $CHR = \frac{n_{ch}}{n_{ch}+n_{cm}}$, where $n_{ch}$, $n_{cm}$ denote number of cache hits and cache misses, respectively. CHR estimates the probability that an incoming read request will be satisfied by cache.

This leads to reduced latency and better overall performance of the system. 

\textbf{Precision.}
There's another important characteristic of the prefetcher algorithms – precision. $Precision = \frac{n_{pr} - n_{eu}}{n_{pr}}$, where $n_{pr}$, $n_{eu}$ denote number of prefetched requests and evicted untouched requests, respectively. It's a measure of how many requests that were uploaded from prefetcher in cache finally led to a cache hit. 

\textbf{Strorage Activity Ratio (SAR).}  
For practical evaluation of prefetching algorithms, we suggest a new metric - storage activity ratio (SAR). This metric estimates of the number of times the Cache with prefetcher system is formed more often to the storage to load blocks into the cache. $SAR = \frac{n_{dp}}{ n_{dc}},$ where $n_{dp}$, $n_{dc}$ denote number of downloaded blocks from storage system cache with prefetcher  and only cache, respectively. This metric characterizes the increase of the storage backend activity due to the prefetcher predictions. 

It should be noted that CHR and precision characteristics contradict each other the more right associations are provided by the prefetcher the higher is CHR in general case but the precision value will have a tendency to decrease. So the challenge for the developers of the prefetching algorithms is to propose such algorithms that would improve both of these characteristics.

Moving on to defining the input parameters of the prefetcher evaluation methodology. 

\textbf{Set of traces.} 
It is a set of I/O traces on each we evaluate prefetching algorithms. It is necessary to find a large enough and relevant dataset to compare the results of prefetching algorithms in conjunction with cache. At the moment the most popular and quoted datasets are: MSR-Cambridge, Tensent and AliCloud.

For evaluation of the read prefetching algorithms, we use MSR-Cambridge traces collection [1] as one of the most cited in storage systems community. This collection contains IO-traces recorded from the 13 servers, spread across 36 volumes with 179 disks in total.  Traces exhibit varying numbers of read and write requests.

\textbf{Cache algorithm.} Least Recently Used (LRU) algorithm is used as the baseline  cache algorithm.

\textbf{Prefetching algorithm.} This paper compares the Mithril, DBSP\_f1, DBSP\_f2 families of prefetching algorithms.

\textbf{Storage system size.} The size of the disc space is selected according to the size of the trace. In particular, size of the disc space is equal   sum of  the maximum address of the request it this trace  and  size of this request. 
In other words, it is considered the maximum usable address in a trace.

\textbf{Cache memory percentage.}  In storage systems, the cache size is usually 1-2 \% of the disc system capacity. So for our measurements we took the value of this coefficient 1 \%. It defines what share of memory from the disc space size is used for the cache and prefetcher.

\textbf{Prefetcher relative size.}  
\begin{figure}
\includegraphics[scale=0.8]{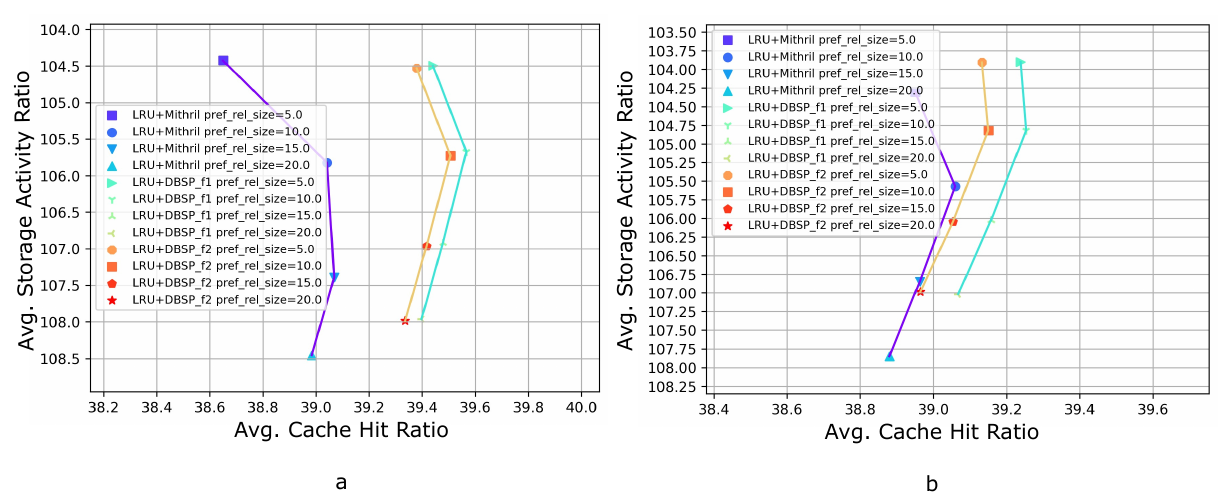} 
\caption{Comparison of prefetching algorithms with different values of prefetcher relative size. (a) Pareto front by Avg. Precision and Avg. Cache Hit Ratio with $s_c = 5$.(b) Pareto front by Avg. Precision and Avg. Cache Hit Ratio with $s_c = 7$. } 
\label{pic:PRS}
\end{figure}
The next input parameter is prefetcher relative size. This parameter defines what part of the available RAM is allocated for the prefetcher. The analysis of the results Figure 2 showed that it is not reasonable to allocate more than 10\% for the prefetcher.

\section{DBSP Evaluation Results}
This chapter presents the experimental results with the DBSP algorithm. The results are presented in the form of two types of graphs.  The  s-curve (see Figure 4) demonstrates the performance of the prefetching algorithms for individual measurements. On the x-axis  is the target metrics (CHR or Precision) , and on the  y-axis of the graph are the trace names from the MSR dataset. Paretto front is used to compare several prefetching algorithms. The x-axis of the graph shows the CHR metric and the y-axis shows the target metrics (Precision or SAR). 

 Results of experiments presented in Figure 3 for algorithms LRU (without prefetching algorithm), LRU+Mithril, LRU+DBSP\_f, LRU+DBSP\_g.
The following parameters were chosen and fixed for the experiments: $l_{min}$ = 2 and $s_{c}$ = \{1, 2, 3, 5, 7, 9\}.

Experimental results show following key findings:
\begin{enumerate}
    \item The DBSP algorithm can increase CHR by 10 \% in average case over plain LRU with negligible additional load on Storage in the order of 3-3.5 \%. 

    \item The DBSP algorithm in average case can increase the CHR by 2\% compared to the SOTA algorithm of Mithril while reducing the additional load on Storage in the order of 1\%. 

    \item The best configuration  of DBSP algorithm has Associated container size equal to $\{2, 3\}$, which demonstrate the maximal increase of CHR and decrease of Precision and SAR.  
    \item An important note is that associated container size greater than 9 does not significantly increase the CHR, while significantly decreasing the accuracy. Therefore, associated container size greater than 9 is not shown in this result.
\end{enumerate}
\begin{figure}
\includegraphics[scale=0.8]{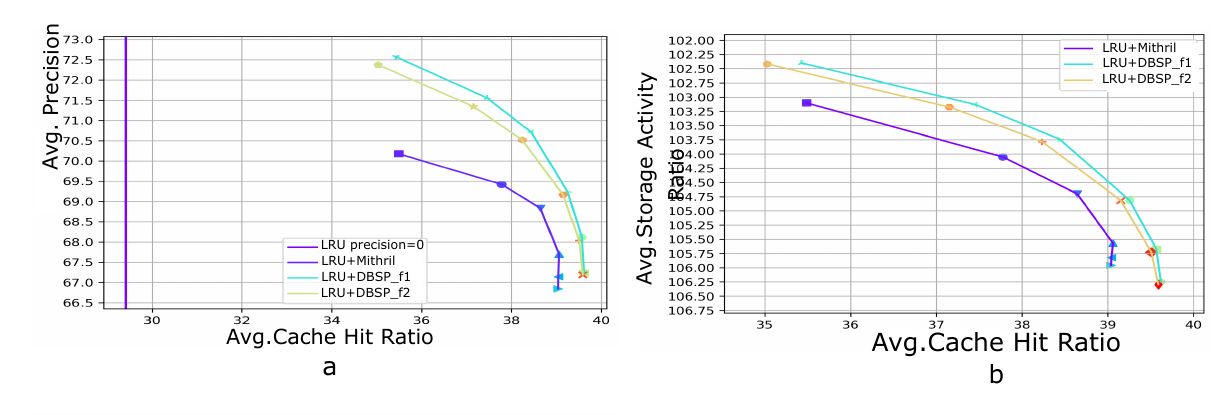} 
\caption{Performance comparison of prefetchers algorithms. (a) Pareto front by Avg. Precision and Avg. Cache Hit Ratio.(b) Pareto front by Avg. Storage Activity Ratio and Avg. Cache Hit Ratio.} 
\end{figure}

\begin{figure}
\includegraphics[scale=0.8]{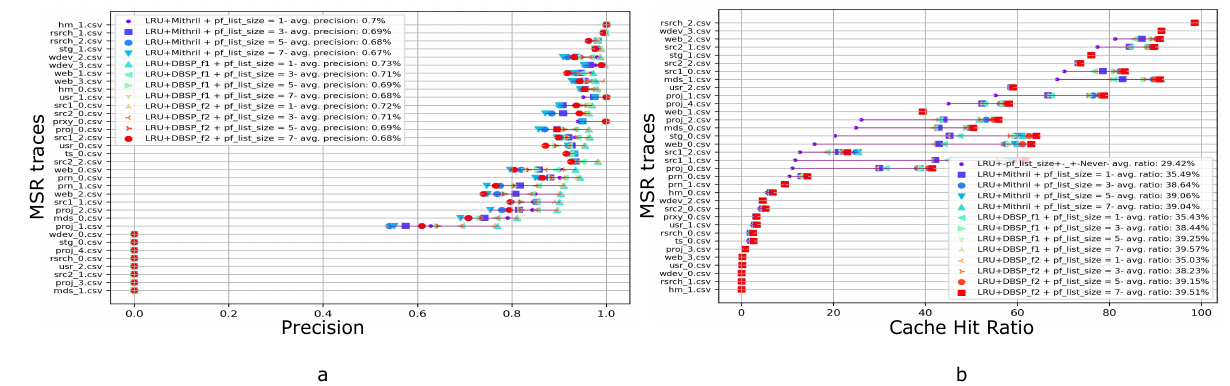} 
\caption{Performance comparison of prefetchers algorithms by traces from MSR dataset. (a) S-curve by  Precision.(b) S-curve Cache Hit Ratio.} 
\end{figure}
The detailed measurement results for each trace separately are shown in Figure 4. Note that for all families of prefetchers  LRU+Mithril, LRU+DBSP\_f1, LRU+DBSP\_f2. Following parameters were chosen and fixed for the experiments: $l_{min}$ = 2 and $s_{c}$ = \{1, 3, 5, 7\}.

\section{Conclusion}
Modern storage systems use read prefetching algorithms to optimize overall performance characteristics. In this paper we propose the new family of prefetching algorithms - Distance Based Sporadic Prefetcher (DBSP). This algorithm family is aimed at discovering sporadic patterns in read request streams. This algorithms overcomes existing state of the art sporadic prefetching algorithm Mithril in the same storage system setup for 2\% in terms of cache hit ratio and 3\% in terms of precision. While it's computational complexity stays almost the same. We also note that DBSP algorithm processes only the history of incoming read requests. It doesn't use any information about the state of the cache or other components of the storage system. This makes it possible to integrate this algorithm into various types of data storage systems. 

Second important result of this paper is the detailed explicit description of the methodology that was used for evaluation of the performance of the read prefetching algorithms. On our experiments this methodology shows consistent results which makes it possible to use it for further investigations in this area.

\bibliographystyle{unsrt}  
% \bibliography{main} 

\begin{thebibliography}{99}
\bibitem[1]{1} Bruce L Jacob. Synchronous dram architectures, organizations, and alternative technologies. University of Maryland,
2002.

\bibitem[2]{2} Seagate. Seagate enterprise capacity hdd. https://www.seagate.com/www-content/datasheets/pdfs/ent-cap-3-5-
hdd-data-sheetds1882-2-1606us-en us.pdf.

\bibitem[3]{3} Seagate. Seagate enterprise capacity ssd. https://www.seagate.com/www-content/datasheets/pdfs/nytro-3000-sasssdds1950-
2-1711gb-en gb.pdf.

\bibitem[4]{4} Binny S. Gill and Luis Angel D. Bathen. AMP: Adaptive multi-stream prefetching in a shared cache. In 5th
USENIX Conference on File and Storage Technologies (FAST 07), San Jose, CA, February 2007. USENIX Association.

\bibitem[5]{5} Jim Griffioen and Randy Appleton. Reducing file system latency using a predictive approach. In USENIX summer,
pages 197–207, 1994.

\bibitem[6]{6} Song Jiang and Xiaodong. Zhang. Lirs: An efficient low inter-reference recency set replacement policy to improve
buffer cache performance. ACM SIGMETRICS Performance Evaluation Review, 30(1):31–42, 2002.

\bibitem[7]{7} Juncheng Yang, Reza Karimi, Trausti Sæmundsson, AvaniWildani, and Ymir Vigfusson. Mithril: Mining sporadic
associations for cache prefetching. 2017.

\bibitem[8]{8} Elizabeth J O’neil, Patrick E O’neil, and Gerhard Weikum. The lru-k page replacement algorithm for database
disk buffering. Acm Sigmod Record, 22(2):297–306, 1993.

\bibitem[9]{9} Dushyanth Narayanan, Austin Donnelly, and Antony Rowstron. Write Off-Loading: Practical power management
for enterprise storage. In 6th USENIX Conference on File and Storage Technologies (FAST 08), San Jose, CA,
February 2008. USENIX Association.


\end{thebibliography}

\end{document}